\documentclass[5p,sort&compress,twocolumn]{elsarticle} 

\usepackage{graphicx}
\usepackage{dcolumn}
\usepackage{bm} 
\usepackage{color}
\usepackage{epstopdf}
\usepackage{amssymb}

\journal{Physics Letters A}

\begin{document} 

\newcommand{\beq}{\begin{equation}}
\newcommand{\eeq}{\end{equation}}
\newcommand{\barr}{\begin{eqnarray}}
\newcommand{\earr}{\end{eqnarray}}

\newcommand{\BLUE}[1]{\textrm{\color{blue}#1}}
\newcommand{\GREEN}[1]{{\color{green}#1}}
\newcommand{\RED}[1]{\textrm{\color{red}#1}}
\newcommand{\rev}[1]{{\color{red}#1}}
\newcommand{\REV}[1] {{\color{red}#1}}
\newcommand{\magenta}[1] {{\color{magenta}#1}}

\newcommand{\andy}[1]{ }
\newcommand{\bmsub}[1]{\mbox{\boldmath\scriptsize $#1$}}

\def\R{\mathbb{R}}

\def\bra#1{\langle #1 |}
\def\ket#1{| #1 \rangle}
\def\sinc{\mathop{\mbox{sinc}}\nolimits}
\def\cV{\mathcal{V}}
\def\cH{\mathcal{H}}
\def\cT{\mathcal{T}}
\def\cM{\mathcal{M}}
\def\cN{\mathcal{N}}
\def\CW{\mathcal{W}}
\def\e{\mathrm{e}}
\def\ii{\mathrm{i}}
\def\d{\mathrm{d}}
\renewcommand{\Re}{\mathop{\mbox{Re}}\nolimits}
\newcommand{\tr}{\mathop{\mbox{Tr}}\nolimits}

\title{Robustness of  raw  quantum tomography }

\author[zar]{M. Asorey} 
\author[barimat,infnba,mec]{P. Facchi} 
\author[barifis,infnba,mec]{G. Florio} 
\author[leb]{V.I. Man'ko\corref{cor1}} 
\ead{manko@lebedev.ru} 
\author[napfis,infnna,mec]{G. Marmo} 
\author[barifis,infnba,mec]{S. Pascazio} 
\author[texfis]{E.C.G. Sudarshan}

\address[zar]{Departamento de F\'\i sica Te\'orica, Facultad de
Ciencias, Universidad de Zaragoza, 50009 Zaragoza, Spain}
\address[barimat]{Dipartimento di Matematica, Universit\`a di Bari,I-70125  Bari, Italy}
\address[infnba]{INFN, Sezione di Bari, I-70126 Bari, Italy}
\address[mec]{MECENAS, Universit\`a Federico II di Napoli \& Universit\`a di Bari, Italy}
\address[barifis]{Dipartimento di Fisica, Universit\`a di Bari, I-70126  Bari, Italy}
\address[leb]{P.N. Lebedev Physical Institute, Leninskii Prospect 53, Moscow 119991, Russia}
\address[napfis]{Dipartimento di Scienze Fisiche, Universit\`a di Napoli ``Federico II", I-80126  Napoli, Italy}
\address[infnna]{INFN, Sezione di Napoli, I-80126  Napoli, Italy}
\address[texfis]{Department of Physics, University of Texas, Austin, Texas 78712, USA}

\cortext[cor1]{Corresponding author}

\begin{abstract}
We scrutinize the effects of non-ideal data acquisition on the tomograms of quantum states. The presence of a weight function, schematizing the
effects of a finite window or equivalently noise, only affects the state reconstruction procedure by a
normalization constant. The results are extended to a discrete mesh and show 
that quantum tomography is robust under incomplete and approximate
knowledge of tomograms.
\end{abstract}


\maketitle

\section{Introduction}

Quantum technological applications require extremely accurate knowledge of
quantum states and of the underlying quantum dynamical processes. For the
application of fundamental principles of quantum mechanics and quantum
optics to effectively foster the real-life implementation of quantum
technologies, accurate quantum state characterization is a crucial ingredient at the interface between theoretical and experimental physics. Applications cover wide research areas ranging from nano-science to cosmology.

One of the most successful quantum state reconstruction techniques is
quantum tomography \cite{Ber-Ber,Vog-Ris} with its elegant experimental
realizations \cite{konst,SBRF93,reconstruct06,pfunction,ps,bellini,allevi09}.  For recent reviews, see \cite{Jardabook,lwry}.  A large class of quantum states, expressed in terms of Wigner functions, can be efficiently
reconstructed by this method, including coherent states, Sch\"odinger cats,
kittles, and entangled states. In quantum optics, the state can be directly 
measured by homodyne photon
detection \cite{Mancini96}, providing as output result the optical tomogram. However, the precise
characterization of quantum states becomes problematic, both for direct measurements and reconstruction
procedures, when noise, imperfections and other practical
problems deteriorate the quality or limit the size of the
data set. This entails, from an experimental perspective, a precise
control/manipulation of the quantum system, from the source to the careful
optimization of the detection apparata, and from a theoretical perspective,
the refinement of mathematical inversion techniques for noisy data and the
extension of classical tomographic techniques to quantum situations \cite{Ber-Ber,Vog-Ris,Mancini96,MarmoPL,banas,hradil,hradilangle}.

In quantum applications, the most used tomographic methods are quantum generalizations of the Radon transform \cite{Rad1917, John, Strichartz}. This is true both for massive particles and reconstruction methods based on
homodyne photon detection. The measure of optical tomograms yields complete information on quantum states in terms of tomographic probability distributions \cite{ibort}. Since, traditionally, one is interested in other equivalent characteristics of quantum states, like the Wigner or the Husimi or the diagonal coherent state representation function, we shall address here the question of the relation between these quantum state characteristics and our approach.

In practice the parameters introduced in the tomogram have many sources of uncertainty
and an efficient tomographic 
measurement of the quantum state must face three major problems:
i) a finite window (including the effects of detectors, entailing coarse graining and/or binning of the  data),
ii) the presence of random errors (arising both from the sample and the
non-ideal precision in controlling the quadratures), and iii) a
discretized ``mesh" in the data acquisition, that affects the generation of
the quadratures of the tomogram.

A realistic approach must define quantum tomograms through a convolution
with a smooth weight function $\Xi $ that spreads the marginals and reduces
to the usual ``classical'' Radon transform when $\Xi $ becomes a delta
function. 
In this case one gets a kind of ``thick tomography," in
analogy with signal analysis, where a similar problem arises for signal
detection and also requires the use of a window function \cite{cohen,gabor}.
Notice that in a classical description, where tomograms are measured by the attenuation of a probing beam scanning a material body, this would correspond to a finite transversal thickness of the scanning beam. However, this is only an analogy: in the physical situation we are describing there is no material body, nor probing beams, and the thickness is in phase space.

Our aim is to suggest extensions of the quantum tomographic techniques by taking into account the finiteness of the windows, the discreteness of data acquisition and the effects of noise, analogously to signal analysis. Previous work \cite{banaszek,artiles} focused on the definition of optimization strategies given a set of experimental data and nonideal photodetector efficiency in homodyne detection (see also \cite{hayashi} for a review).

This paper is organized as follows. In Sec.\ \ref{sec:prel} we recall the main concepts related to the Radon and symplectic transforms. Section \ref{sec:finite} is devoted to the effects of the finite window function. In Sec.\ \ref{sec:noise} and \ref{sec:discrete} we analyze how the tomograms are modified by noise and a discrete mesh.
In Sec.\ \ref{sec:robust} we present some numerical results for a particular example, corroborating our analysis.
Finally, in Sec.\ \ref{sec:concl} we sketch some conclusions and outline possible future research directions.

\section{Preliminaries: Radon and symplectic transforms}
\label{sec:prel} 

We shall start by studying the role of the deformation associated with the window weight
function in the experimental tomogram and its influence on the state reconstruction
formula. The following framework is of general validity, and can be applied to massive particles as well as photons.
However, for simplicity, let $\widehat{\rho}$ be a given quantum photon state, $\widehat{q}$ and $\widehat{p}$ the position and momentum operators,
and $\varphi $ and $X$ the local oscillator phase and quadrature in a homodyne experimental setup.
The homodyne tomogram is given by
\begin{equation}
\CW(X,\varphi) =\tr \left\{\widehat{\rho}\ \delta\left(X -\widehat{q} \cos\varphi - \widehat{p} \sin \varphi \right) \right\},
 \label{homodyne}
\end{equation}
where $\tr\{A\}$ denotes the trace of the operator $A$.
Expressed in terms of the Wigner function  
\beq
W(p,q)=\frac{1}{\pi }\int_{\mathbb{R}} \langle q-\xi |\widehat{\rho}
|\,q+\xi  \rangle\, \e^{2\ii p\xi}\; \d\xi ,
\eeq
the  tomogram (\ref{homodyne}) reads
\barr
\label{eq:radon}
\mathcal{W}(X,\varphi )&=& \int_{\mathbb{R}^2}\d p\,\d q\; W(p,q)\nonumber\\ &&\times\delta
\left( X-q \cos \varphi - p \sin \varphi \right).
\earr
The Radon transform of the Wigner function (\ref{eq:radon}) has been generalized to the following symplectic, or $M^2$, transform \cite{MarmoPhysScr,MarmoPL}
\beq
\label{eq:m2}
\mathcal{W}^\sharp(X,\mu,\nu )= \int_{\mathbb{R}^2} W(p,q)\, \delta
\left( X-q \mu - p \nu \right)\; \d p\,\d q ,
\eeq
with $\mu, \nu \in \mathbb{R}$. Its complete equivalence with (\ref{eq:radon}) is expressed by the relation \cite{m2,tomolectures}
\begin{equation}
\mathcal{W}^\sharp(X, r \cos\varphi , r \sin \varphi) = \frac{1}{r}\, \mathcal{W}\left(\frac{X}{r},\varphi\right),  
\label{eq:m2radon}
\end{equation}
valid for any $r>0$. Equation~(\ref{eq:m2radon}) is an easy consequence of the fact that the Dirac distribution is positive homogeneous of degree $-1$. 
The symplectic tomogram  (\ref{eq:m2}) can be easily inverted by a Fourier transform
\begin{equation}
W(p,q)= \int_{\mathbb{R}^3} \frac{\d X\,\d\mu\,\d\nu}{{(2\pi)}^2}\;  \mathcal{W}^\sharp(X,\mu,\nu )\, \e^{\ii (X- q \mu - p\nu)},
\end{equation}
which by (\ref{eq:m2radon}) yields the inversion formula
\begin{eqnarray}
W({p},{q}) &=&
\frac{1}{(2\pi)^2 }
\int_{\mathbb{R}} \d X \int_ {\mathbb{R}_+}  \d r \int_ {\mathbb{T}} \d \varphi\; \mathcal{W}(X,{\varphi })
\nonumber \\
&&\times r\, \e^{\ii r(X-{q}\cos \varphi -{p}\,{\sin \varphi })},
 \label{im2sim}
\end{eqnarray}
where $\mathbb{T}= \mathbb{R}/2\pi \mathbb{Z}$ is the unit circle and $\mathbb{R}_+=[0,+\infty)$.

\section{Finite window}
\label{sec:finite}

A finite window function can be accounted for by replacing the
Dirac delta function in the definition of the tomogram (\ref{eq:m2}) with a suitable smearing window function $\Xi$
\beq
\label{eq:m2thick}
\mathcal{W}^\sharp_{\Xi }(X,\mu,\nu )= \int_{\mathbb{R}^2} W(p,q)\, \Xi
\left( X-q \mu - p \nu \right) \d p\,\d q .
\eeq
Now, by Fourier transforming Eq.~(\ref{eq:m2thick}), one gets
\barr
W(p,q)&=& \mathcal{N}_{\Xi} \int_{\mathbb{R}^3} \frac{\d X\,\d\mu\,\d\nu}{{(2\pi)}^2}\;  \mathcal{W}^\sharp_{\Xi }(X,\mu,\nu )\nonumber\\ &\times&\e^{\ii (X- q \mu - p\nu)},
\label{iradonsim}
\earr
where
\begin{equation}
\mathcal{N}_{\Xi }=\frac{1}{\widetilde{\Xi }(-1)},\quad \widetilde{\Xi }
(-1)= \int_ {\mathbb{R}}  {\Xi }(z)\, \e^{\ii z}\; \d z .
\label{formula1}
\end{equation}
In operator terms the state reconstruction is achieved by 
\begin{equation}
\widehat{\rho } =
\mathcal{N}_{\Xi} \int_{\mathbb{R}^3} \frac{\d X\,\d\mu\,\d\nu}{{(2\pi)}^2}\;  \mathcal{W}^\sharp_{\Xi }(X,\mu,\nu )\, \e^{\ii (X- \widehat{q} \mu - \widehat{p}\nu)}.
\label{formulaa}
\end{equation}
This is one of our central results: the inverse reconstruction map is \emph{independent} of the window
function, the only relic of the latter being the normalization
constant $\cN_\Xi$, that can be fixed by the normalization of one single tomogram.
Thus, no matter how involved is the shape of the window function, an exact
reconstruction of the state can be always achieved by means of (\ref
{iradonsim}) or (\ref{formulaa}). 
This property is very interesting because in practical cases the experimental
window function $\Xi$ is unknown, but the result tells us that the exact state reconstruction is
possible without any precise information about $\Xi$.
This result is also independent of the features of the initial state,
that can even be nonstationary. Interestingly, this result can be \emph{extended} to situations were noise is present. This will be done in the next section.

\section{Noise}
\label{sec:noise}

We start from the following important observation:
the effects of a spread function in the tomogram (\ref{eq:m2thick})
are \emph{de facto} equivalent to the presence of noise, due to quantum
efficiency and/or thermalization, in a homodyne tomogram (e.g., if the quadrature is determined with a finite
precision). 
The origin of such noise is due to the fact that any counting homodyne statististics is just a sampling of probabilities.
Namely, any statistical event can be only approximately sampled and in the course of repeated measurements, the outputs are always fluctuating. All this can be seen as a jitter in the detected statistics.

The equivalence between such a jitter and a finite window can be seen as follows: if there is some jitter $X+\xi$ on $X$, where $\xi$ is a
random variable with a finite variance, one obtains 
\begin{equation}
\mathcal{W}^{(\xi)}(X,\mu,\nu) =\mathrm{Tr}\left\{\widehat{\rho}\ \delta\left(X +\xi
-\widehat{q} \mu - \widehat{p} \nu \right)\right\}.
\label{shomodynenoise}
\end{equation}
By setting
\begin{equation}
\mathbb{E}[\delta\left(X +\xi\right)]= \Xi(X), 
 \label{ee0}
\end{equation} 
where $\mathbb{E}$ denotes the expectation value over the random variable $\xi$,
one gets 
\begin{equation}
\mathbb{E}[\mathcal{W}^{(\xi)}(X,\mu,\nu)] = \mathcal{W}^\sharp_\Xi(X,\mu,\nu) 
 \label{ee}
\end{equation} 
and the reconstruction can be done exactly as above.

The jitter provides extra noise and deforms the
output of homodyne experiments by affecting the basic inequalities of quantum mechanics expressed in
terms of tomograms. For this reason, it is of practical interest to take it into account when extracting 
information on the uncertainty relations from the measurements of 
tomograms. In practice this can be done, for example, by using the output of photon homodyne
measurements \cite{SBRF93,lwry}, given as optical tomograms (\ref{homodyne}).
The quadrature (co)variances are directly expressed in terms
of simple integrals containing optical tomograms \cite{ibort,mankonew} and can be used to yield a noise limit contribution to the experimental accuracy of
uncertainty relations in homodyne photon detection.

Let us look at a simple but significant example of experimental relevance. Consider 
the Wigner function of the first excited state of a harmonic oscillator (or a single photon state) 
\beq
W(p,q)=\frac{2(p^2+q^2)-1}{\pi} \, \e^{-(p^2+q^2).}
\label{eq:wignersinglephoton}
\eeq
Its Radon transform reads
\beq
\mathcal{W}(X,\varphi)=  \frac{2 X^2}{\sqrt{\pi}  }\,   \e^{-X^2}.
\label{eq:radonsinglephoton}
\eeq
The introduction of a Gaussian window function 
\begin{equation}
\Xi(X)=\frac{1}{\sqrt{2\pi\sigma^2}} \, \e^{-X^2/ 2\sigma^2},
\label{eq:windowgauss}
\end{equation}
yields the symplectic tomograms
\begin{eqnarray}
\mathcal{W}^\sharp_\Xi(X,\mu,\nu ) &=& \frac{2 (X^2/r^2+\sigma^2(2\sigma^2+1)) }{\sqrt{\pi} r(2\sigma^2+1)^{5/2}}\nonumber\\  &&\times \e^{-X^2/r^2(2\sigma^2+1)},
\label{eq:smoothed}
\end{eqnarray}
with $r=\sqrt{\mu^2+\nu^2}$.
Note that 
\begin{equation}
\mathcal{W}^\sharp_\Xi(0,\mu,\nu)=  \frac{2 \sigma^2}{\sqrt{\pi r^2 (2\sigma^2+1)^{3}}}. 
\end{equation}
Therefore, the presence of a window function/noise provokes a reduction of ``visibility" of the tomogram (that would vanish for $\sigma=0$). 
However, a perfect state
reconstruction is obtained by Eqs.\ (\ref
{iradonsim}) or (\ref{formulaa}). 

This proves that an \emph{exact} reconstruction can be obtained even in the presence of noise, provided one has very (in the limit, infinitely) accurate control over the position of the quadrature. This extends the central result of the preceding section.

\section{Discreteness of data acquisition}
\label{sec:discrete} 

So far we have assumed that the window function can be finite (thick tomograms), but one has access to all possible tomograms in a continuous range of
parameters.
We have not discussed the robustness of tomography
with respect to the discreteness of data acquisition.
Let us assume that the experimental tomograms are gathered only on a
sequence of discrete values of $X$ and $\varphi$, namely 
\begin{equation}
\mathcal{W}_{k, m} =\mathcal{W}\left(k T, m \frac{2 \pi }{N} \right), \qquad
k\in \mathbb{Z}, \; m\in\mathbb{Z}_N  \label{eq:samples}
\end{equation}
where $T>0$, $\mathbb{Z}_N=\mathbb{Z}/N\mathbb{Z}$ and, for convenience, $N$ is an
odd positive integer. If the tomograms have a limited bandwidth, then for
sufficiently small values of $T$ and $N^{-1}$ one can exactly reconstruct the
whole family of tomograms from the knowledge of the experimental ones. This
is the content of the Nyquist-Shannon sampling theorem \cite
{Nyquist,Shannon}. More precisely, consider the Fourier transform of $\mathcal{W}$
\begin{equation}\label{eq:tildeW}
\widetilde{\mathcal{W}}(\omega,\ell)=\int_{\mathbb{R}} \d X
\; \e^{-\ii \omega X} \int_{\mathbb{T}}\frac{\d \varphi}{2\pi}\; 
\e^{-\ii \ell \varphi}\, \mathcal{W}(X,\varphi).
\end{equation}
If its support is
compact and satisfies
\begin{equation}
\mathrm{supp}\,\widetilde{\mathcal{W}} \subset D, \quad D= \Big( -\frac{\pi}{%
T},\frac{\pi}{T}\Big)\times \Big( -\frac{N}{2},\frac{N}{2}\Big),
\label{eq:samp}
\end{equation}
one has 
\barr\label{eq:discretetildeW}
\widetilde{\mathcal{W}}(\omega,\ell)&=&\frac{T}{N}\sum_{k\in\mathbb{Z}}\sum_{m\in%
\mathbb{Z}_N} \mathcal{W}_{k, m}\, \e^{-\ii X_k \omega} \, \e^{-\ii \varphi_m 
\ell} \nonumber\\
&&\times\chi_D(\omega,\ell),
\earr
where $X_k=kT$, $\varphi_m=2\pi m/N$, and

\begin{equation}
\chi_D(\omega,\ell)=
\left\{
\begin{array}{rl}
 {1} & {\mbox{if}\quad (\omega, l)\in D }\\
 {0} & {\mbox{if}\quad  (\omega, l)\notin D.}
\end{array}
\right.
\end{equation}

Let us briefly show how (\ref{eq:discretetildeW}) is obtained, by considering the less common situation of a function on the torus $\mathbb{T}$:
\begin{equation}
\widetilde{f}(\ell)= \int_{\mathbb{T}}\frac{\d \varphi}{2\pi}\; 
\e^{-\ii \ell \varphi}\, f(\varphi),
\quad
f(\varphi)=\sum_{\ell\in\mathbb{Z}} \e^{\ii \ell \varphi}\, \widetilde{f}(\ell).
\end{equation}
Suppose $\widetilde{f}(\ell)=0$ for $\ell\notin J$, with $J=(-N/2,N/2)\cap \mathbb{N}$. For definiteness let us assume $N$ odd.
Then $J=\{-(N-1)/2,\dots, (N-1)/2\}$, and one can consider the periodic extension of $\widetilde{f}$
\begin{equation}
\widetilde{f}_N(\ell)= \sum_{k\in\mathbb{Z}} \widetilde{f}(\ell - k N),
\end{equation}
whose restriction to $J$ coincides with $\widetilde{f}$, namely,
\begin{equation}
\widetilde{f}(\ell) =\widetilde{f}_N(\ell)\,\chi_J(\ell).
\label{eq:noalias}
\end{equation}
This property would not be true if $\widetilde{f}$ did not vanish outside $J$, and the phenomenon called aliasing would occur. One gets
\begin{equation}
\widetilde{f}_N(\ell) = 
\int_{\mathbb{T}}\frac{\d \varphi}{2\pi}\; 
\e^{-\ii \ell \varphi}\, f(\varphi)
 \sum_{k\in\mathbb{Z}} \e^{\ii k N \varphi}.
\end{equation} 
By recalling Poisson's formula
\begin{equation}
 \sum_{k\in\mathbb{Z}} \e^{\ii k N \varphi}
= \frac{2\pi}{N} \sum_{k\in\mathbb{Z}}\delta\left(\varphi-\varphi_k \right),
\label{eq:Poisson}
\end{equation}
with $\varphi_k=2 k\pi/N$,
and setting $k= j N + m$ with $k, j \in \mathbb{Z}$ and $m\in\mathbb{Z}_N$ one gets
\begin{equation}
 \sum_{k\in\mathbb{Z}} \e^{\ii k N \varphi}
= \frac{2\pi}{N} \sum_{m\in\mathbb{Z}_N} \sum_{j\in\mathbb{Z}}\delta\left(\varphi-\varphi_m -2 j \pi\right).
\end{equation}
Therefore,
\begin{equation}
\widetilde{f}_N(\ell) = \frac{1}{N}
 \sum_{m\in\mathbb{Z}_N} 
\e^{-\ii \ell \varphi_m}\, f(\varphi_m).
\label{eq:tildefN}
\end{equation}
By plugging (\ref{eq:tildefN}) into (\ref{eq:noalias}) one gets the angular dependence of (\ref{eq:discretetildeW}). The linear dependence is obtained analogously by  replacing $\mathbb{T}$ with $\mathbb{R}$ and $\mathbb{Z}_N$ with $\mathbb{Z}$.

By Fourier inverting (\ref{eq:noalias}) one gets
\begin{eqnarray}
f(\varphi) & = & \sum_{\ell\in J} \e^{\ii \ell \varphi}\, \widetilde{f}_N(\ell)\nonumber\\
&=& \sum_{m\in\mathbb{Z}_N} f(\varphi_m)
\frac{1}{N}\sum_{\ell\in J} \e^{\ii \ell (\varphi-\varphi_m)} \nonumber\\
& = & \sum_{m\in\mathbb{Z}_N} f(\varphi_m)\,
S_N \left(\frac{\varphi-\varphi_m}{2}\right),
\label{ eq:FSN}
\end{eqnarray}
with
\begin{equation}
\label{ eq:SNdef}
S_N(x)=\frac{\sin (N x)}{N \sin x},
\end{equation}
which is the extension of the sampling theorem to functions on the torus $\mathbb{T}$ and their discrete spectra. For a function $g(X)$ on the line, the analogous, well-know formula reads
\begin{equation}
g(X) = \sum_{k\in\mathbb{Z}} g(X_k)\,
\sinc\left(\pi \frac{X-X_k}{T}\right), 
\end{equation}
with $X_k=k T$ and $\sinc x = x^{-1} \sin x$.

Therefore, the tomograms are given, for $N$ odd, by the (generalized) Shannon-Whittaker
interpolation formula 
\begin{eqnarray}
\CW\left(X,\varphi \right)&=& \sum_{k\in\mathbb{Z}}\sum_{m\in\mathbb{Z}_N}
\CW_{k, m}\, \sinc\left(\pi \frac{X-X_k}{T}\right)\nonumber\\ &&\times S_N\! \left(\frac{\varphi-\varphi_m}{2}\right),  
\; \forall X \in\mathbb{R},  \forall \varphi\in\mathbb{T}.\nonumber\\
\label{eq:SW}
\end{eqnarray}
For $N$ even the formula is the same, by replacing $N$ by $N-1$ (since in that case the angular sampling points are $N-1$ instead of $N$). Remarkably, for sufficiently small $T$ and $N^{-1}$ the reconstruction of limited-bandwidth tomograms is \emph{faithful} and there is \emph{no} information loss.
It is interesting to observe that the application of signal-analysis techniques to quantum tomography can be quite straightforward. For example, from the mathematical point of view, the extension of Eqs.\ (\ref{ eq:FSN}) and (\ref{eq:SW}) of quantum tomograms to a torus is very natural.

By plugging (\ref{eq:SW}) into (\ref{im2sim})  one can obtain a reconstruction formula of the Wigner function in terms of the tomographic samples (\ref{eq:samples}). By writing
\begin{equation}
\sinc\left(\pi \frac{X-X_k}{T}\right)=\frac{T}{2\pi} \int_{-\pi/T}^ {+\pi/T}  \e^{\ii\omega(X-X_k)}\; \d\omega 
\end{equation} 
and by performing the integration over $X$ one gets a Dirac delta function $\delta(\omega+r)$ whose integral yields 
\begin{eqnarray}
W(p,q) &=&
\sum_{k\in\mathbb{Z}}\sum_{m\in\mathbb{Z}_N}
\CW_{k, m}
 \int_ {\mathbb{T}} \d \varphi \; S_N\! \left(\frac{\varphi-\varphi_m}{2}\right)
\nonumber \\
&&  \times \frac{T} {(2\pi)^2 }\int_0^{\pi/T}  \d r 
r \e^{\ii r(X_k-{q}\cos \varphi -{p}\,{\sin \varphi })}.\nonumber\\
\end{eqnarray}
We get
\begin{eqnarray}
\frac{T} {(2\pi)^2 }\int_0^{\pi/T}  \d r \;
r\, \e^{\ii r\beta} &=& \frac{\alpha\sin\alpha + \cos\alpha - 1}{4 T \alpha^2}
\nonumber\\
& &+ \ii \frac{\alpha\cos\alpha-\sin\alpha} {4 T \alpha^2},
\end{eqnarray}
with $\alpha=\pi \beta /T$.
Since the imaginary part is odd with respect to the rotation $\varphi \to \varphi + \pi$, its integral over the torus vanishes and we finally obtain
\barr
\label{eq:reconstruction}
W(p,q) &=&
\sum_{k\in\mathbb{Z}}\sum_{m\in\mathbb{Z}_N}\CW_{k, m}
\int_{\mathbb{T}} \d \varphi S_N\! \left(\frac{\varphi-\varphi_m}{2}\right)\nonumber\\
&&\times \left[  \frac{\cos(\alpha_{k}(\varphi;q,p))-1}{4 T \alpha_{k}(\varphi;q,p)^2}\right.\nonumber\\
&&\left.+\frac{\alpha_{k}(\varphi;q,p)\sin(\alpha_{k}(\varphi;q,p))}{4 T \alpha_{k}(\varphi;q,p)^2}\right]
,
\earr
with $\alpha_{k}(\varphi;q,p)=\pi (X_k- q\cos\varphi - p\sin\varphi)/T$ (and $N$ odd).

Let us now consider an experimental situation in which there is some
uncertainty on the linear and angular position of the quadrature, that is 
\begin{eqnarray}
& &\mathcal{W}_{k, m}^{(\xi)} =\mathcal{W}\left( X_k^{(\xi)} , \varphi_m^{(\xi)} \right), 
\nonumber \\
& & X_k^{(\xi)}= T(k +\xi_{k}^{(1)}), \quad \varphi_m^{(\xi)} =  \frac{2 \pi }{N}(m+\xi_{m}^{(2)}) ,
\label{eq:noises}
\end{eqnarray}
with $k\in \mathbb{Z}$ and $m\in\mathbb{Z}_N$, where $\{\xi_{k}^{(1)}\}$ and $\{\xi_{m}^{(2)}\}$ are two sequences of independent identically distributed
random variables with finite standard deviations $\sigma^{(1)}_\xi$ and $\sigma^{(2)}_\xi$. Under hypothesis (\ref{eq:samp}), one
gets 
\barr
\widetilde{\mathcal{W}}(\omega,\ell)&=&\frac{T}{N} \mathbb{E}\Big[ \sum_{k\in\mathbb{Z}}\sum_{m\in\mathbb{Z}_N}\!\! \mathcal{W}_{k, m}^{(\xi)}\, \e^{-\ii X_k^{(\xi)}
\omega}\, \e^{-\ii \varphi_m^{(\xi)} \ell} \nonumber\\ && \times\chi_D(\omega,\ell) \Big].
\label{eq:average}
\earr
A Fourier transform, followed by a Radon inversion (\ref{formulaa}) yields on average a \emph{perfect} reconstruction of the state. Therefore, quadrature uncertainties and unbiased noise do \emph{not} affect the result.

Let us prove Eq.~(\ref{eq:average}). Consider a function $g(X)$ on the line, with limited bandwidth, namely, $\widetilde{g}(\omega)=0$ for $\omega\notin I$, where $I=(-\pi/T,+\pi/T)$. One can write
\begin{eqnarray}
h^{(\xi)}(\omega)&=&\sum_{k\in\mathbb{Z}} g(X_k^{(\xi)}) \, \e^{-\ii X_k^{(\xi)}\omega}
\nonumber\\
& = &
\sum_{k\in\mathbb{Z}}  \int \frac{\d\nu}{2\pi} \; \widetilde{g}(\nu) \, \e^{-\ii X_k^{(\xi)}(\omega-\nu)}
 \nonumber\\
& = &
 \int \frac{\d\nu}{2\pi} \; \widetilde{g}(\omega-\nu) \, \sum_{k\in\mathbb{Z}} \e^{-\ii X_k^{(\xi)}\nu}
\end{eqnarray}
By taking the expectation value, and by using Poisson's formula (\ref{eq:Poisson}) one gets
\begin{eqnarray}
\mathbb{E}\Big[\sum_{k\in\mathbb{Z}} \e^{-\ii X_k^{(\xi)}\nu}\Big]
&=& \mathbb{E} \left[\e^{-\ii \xi \nu}\right] \sum_{k\in\mathbb{Z}} \e^{-\ii X_k \nu}
\nonumber\\
&=& \frac{2\pi}{T} \mathbb{E} \left[\e^{-\ii \xi \nu}\right] \sum_{k\in\mathbb{Z}} \delta\left(\nu-\frac{2\pi k}{T}\right). \nonumber\\
\end{eqnarray}
Therefore,
\barr
\mathbb{E}\left[h^{(\xi)}(\omega) \right] 
&= &\frac{1}{T} \sum_{k\in\mathbb{Z}}  \widetilde{g}\left(\omega-\frac{2\pi k}{T}\right)\mathbb{E} \left[\e^{-\ii 2\pi k \xi/ T}\right],\nonumber\\
\earr
that, when multiplied by $\chi_I(\omega)$, yields
\begin{equation}
\mathbb{E}\left[h^{(\xi)}(\omega) \chi_I(\omega) \right]
= \frac{1}{T}\,  \widetilde{g}\left(\omega\right)\mathbb{E} \left[1\right] ,
\end{equation}
because of the condition on the bandwidth of $g$. Since $\mathbb{E} \left[1\right]=1$, we get the linear dependence of (\ref{eq:average}), the proof of the angular dependence being analogous.

A comment seems in order. In our analysis 
we separately addressed the effects that arise from the existence of a finite window (or equivalently the presence of noise) and those that are a consequence of partial data acquisition. 
The problems that arise in non-ideal situations are under control in two limiting cases: when the window size $\sigma$ and the sampling steps $s(=T, 1/N)$ are well separated, i.e.\ $\sigma\ll s$ or $s\ll \sigma$. The first case can be analyzed through Eq.\ (\ref{eq:reconstruction}), the second one through Eqs.\ (\ref{iradonsim})-(\ref{formula1}). It would be interesting to understand whether a single general formula exists, that accounts at the same time for the consequences of all these effects and from which both above-mentioned cases arise as suitable limits. 
Such a formula would also enable us to elucidate whether the simple expression we obtained in terms of the normalization constant
(\ref{formula1}) can be generalized to the case of a finite number of measurements, or 
when noise is combined with other sources of uncertainty. This interesting aspect is left for future research. The following section is devoted to a partial numerical investigation of this problem.

\section{Robustness of tomograms}
\label{sec:robust} 

So far, our analysis has taken into account the effects of noise and discreteness of data. We have seen that for sufficiently small $T$ and $N^{-1}$ the reconstruction of limited-bandwidth tomograms is faithful and there is no information loss [see Eq.\  (\ref{eq:SW}) and following comments]. Also, quadrature uncertainties and unbiased noise do not affect the reconstruction [see Eq.\  (\ref{eq:average}) and following comments].
By combining all these results we now discuss the robustness of quantum tomograms under the afore-mentioned 
sources of uncertainty and limitations (discreteness of the sampling and finite precision in the determination of the quadrature, since we have already argued that a finite window is equivalent to the presence of noise). 
Let us look again at the tomogram (\ref{eq:radonsinglephoton}) and  numerically investigate the effects that arise due to a discrete mesh and the presence of noise on the Wigner function (\ref{eq:wignersinglephoton}).
We shall reconstruct the Wigner function by using Eq.~(\ref{eq:reconstruction}) with $T=0.1$, $N=5$, and  $\sum_{k\in\mathbb{Z}}$ replaced by  $\sum_{|k| \leq K}$, with $K=40$.

\begin{figure}
\includegraphics[width=0.4\textwidth]{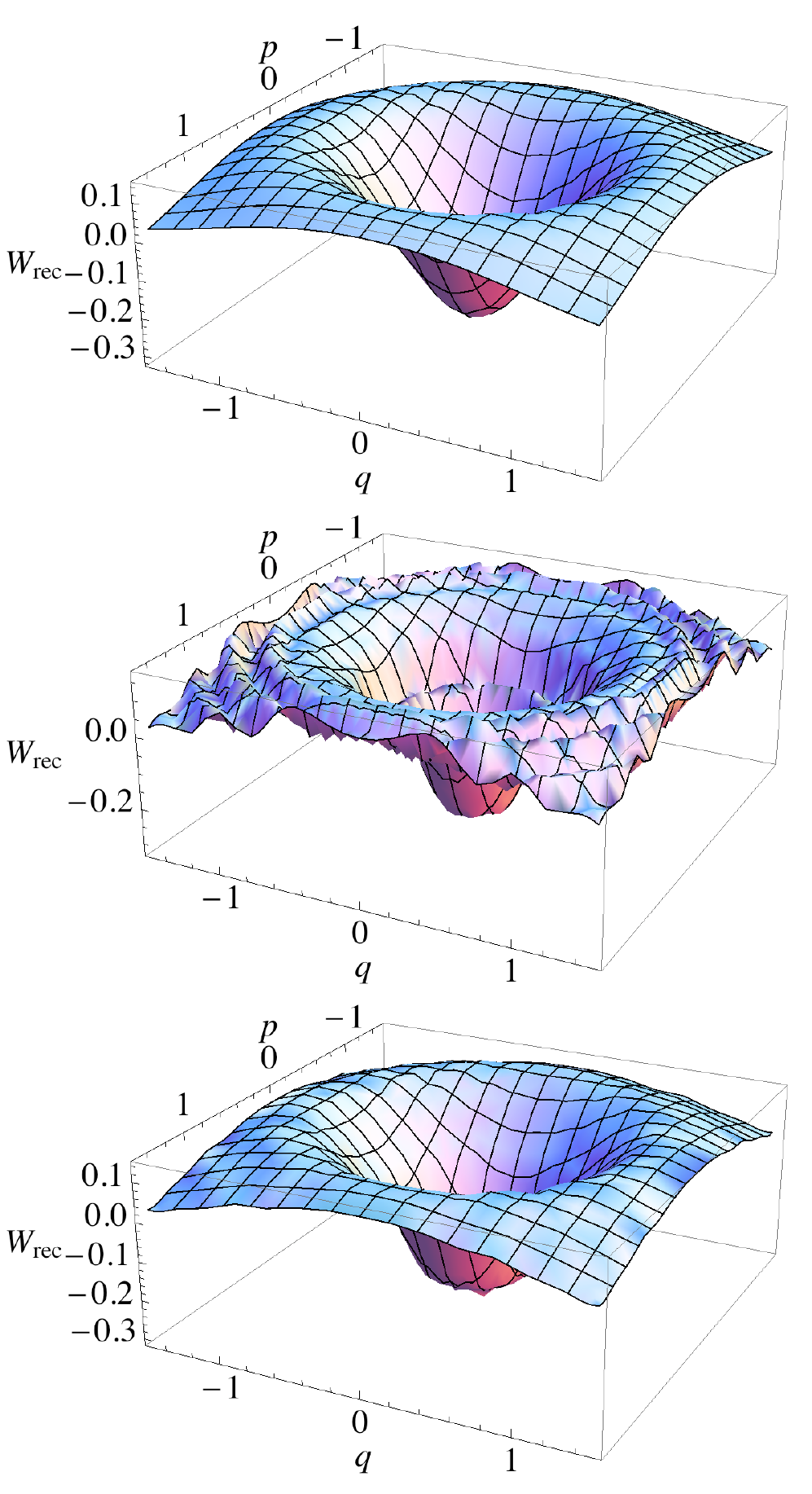}
\caption{Reconstruction starting from the tomograms (\ref{eq:radonsinglephoton})
of the Wigner function $W(p,q)$ in Eq.\ (\ref{eq:wignersinglephoton}). 
We set $N=5$, $T=0.1$ and $K=40$ in Eqs.\ (\ref{eq:reconstruction}) and (\ref{eq:average}).
Upper panel: no noise. The reconstructed and the original function are indistinguishable. Central panel:
single noise realization, $\sigma^{(1)}_\xi=\sigma^{(2)}_\xi=0.5$. Lower panel: average over $10$ realizations of the noise.}
\label{fig:confronto}
\end{figure}

Figure \ref{fig:confronto} shows the tomographic reconstructions of the Wigner function (\ref{eq:wignersinglephoton}), starting from its tomograms (\ref{eq:radonsinglephoton}).
The Wigner function in the noiseless case 
is shown in the upper panel and is practically indistinguishable from the original (\ref{eq:wignersinglephoton}).
Notice that, even though the tomogram  is not band limited and the number of points $K$ is finite, the reconstruction by means of (\ref{eq:reconstruction})  is practically alias free.
The reconstructions in the central and lower panel are affected by noise, as in (\ref{eq:noises}) with zero means and
 $\sigma^{(1)}_\xi=\sigma^{(2)}_\xi=0.5$: these are large values, of the same order of the sampling periods $T$ and $2\pi/N$
(one should notice, however, that in this particular case the noise on the angular position of the quadrature does not affect the procedure due to the symmetry of the state considered). The central panel refers to a single realization of the noise, while the lower panel to an average over 
$10$ realizations of the noise.

The reconstruction error is measured by the distance
\begin{eqnarray}
\varepsilon & = & \frac{1}{2} ||W - W_{\rm rec}||_B \nonumber \\
& = & \frac{1}{2} \int_{B}  |W(p,q)-W_{\rm rec}(p,q) |\; \d p\, \d q ,
\label{eq:discreto}
\end{eqnarray}
where $B$ is the box considered (in our case $B=[-1.5,1.5]^2$). 
We obtain $\varepsilon= 1.4 \times 10^{-4}, 0.11, 0.028$ for the upper, central and lower panel, respectively. This should be compared with $\|W\|_B=1.01$. It is apparent both from Fig.\ \ref{fig:confronto} and the above numerical results that the reconstruction is robust \emph{at the same time} against noise and discretization effects.

\section{Conclusions}
\label{sec:concl}

In conclusion, we have discussed the practical problems that arise in quantum (homodyne) tomography.
First of all, the presence of a weight function has been shown to introduce only a normalization constant, which entails no loss of information and permits the exact reconstruction without any precise knowledge of the window function of the experimental setup: on average, quadrature uncertainties and unbiased noise do not affect the reconstruction.
Second, the discretization procedure affects the global reconstruction in the same way as it does in the classical  Nyquist-Shannon setting: if the bandwidth of the tomogram is limited, there is no information loss for a sufficiently dense sampling. We have also discussed the most general case, in which both problems arise simultaneously. Although we were not able to derive a general formula, from which both sources of non-ideal data acquisitions are present at the same time, and from which both situations investigated here are derived as limiting cases, we proved by numerical methods
the robustness of the overall reconstruction against the different sources of imperfections. The validity of the approach proposed here should be tested in conjunction with other refined theoretical tools based on the maximum likelihood estimation \cite{banas,hradil,banaszek}.

Since nowadays quantum states reconstruction is based on the measurement of tomograms, the robustness against non-ideal data acquisition provides solid foundation to the 
compatibility of tomographic experiments performed in different laboratories by different methods.
Our results provide therefore a solid basis for the theoretical and phenomenological analysis of real tomograms.

\section{Acknowledgments}
V.I.M.\ was partially supported by the Russian
Foundation for Basic Research under Project No. 09-02-00142 and Italian INFN and thanks the Physics
Department of the University of Naples for the kind hospitality. 
P.F.\ and G.F.\ acknowledge support through the project IDEA of University of Bari.
The work of M.A.\ and G.M.\ was partially
supported by a cooperation grant INFN-CICYT. M.A. was also partially
supported by the Spanish CICYT grant FPA2009-09638 and DGIID-DGA (grant 2009-E24/2).


\end{document}